\documentclass[epj-spec-arxiv]{svjour-arxiv}
\usepackage{graphics}
\usepackage{subeqn}
\usepackage{epsf}
\begin{document}
\title{Should particle trajectories comply with the transverse momentum
distribution?}
\author{Milena Davidovi\'c\inst{1} \and Du\v san Arsenovi\'c\inst{2}
\and Mirjana Bo\v zi\'c\inst{2} \and Angel S. Sanz\inst{3}
\and Salvador Miret-Art\'es\inst{3}}
\institute{%
Faculty of Civil Engineering, Bulevar Kralja Aleksandra 73,
1100 Belgrade, Serbia \\
 \email{milena@grf.bg.ac.yu}
\and
Institute of Physics, Pregrevica 118, 11080 Belgrade, Serbia \\
 \email{arsenovic@phy.bg.ac.yu;bozic@phy.bg.ac.yu}
\and
Instituto de Matem\'aticas y F\'{\i}sica Fundamental,
Consejo Superior de Investigaciones Cient\'{\i}ficas,
Serrano 123, 28006 Madrid, Spain \\
 \email{asanz@imaff.cfmac.csic.es;s.miret@imaff.cfmac.csic.es}
}
\abstract{%
The momentum distributions associated with both the wave function of a
particle behind a grating and the corresponding Bohmian trajectories
are investigated and compared. Near the grating, it is observed that
the former does not depend on the distance from the grating, while the
latter changes with this distance. However, as one moves further apart
from the grating, in the far field, both distributions become
identical.
} 
\maketitle
\section{Introduction}
\label{intro}

Quantum interference experiments where beams of one particle at a time
are used have intensified the theoretical investigation of the topology
of particle trajectories behind the interference grating \cite{ref1,%
ref2,ref3,ref4,ref5,ref6,ref7,ref8}. The aim
of all the approaches developed in this direction is to get consistency
between the quantum mechanical particle distribution and the
distribution associated with the trajectories of the particles. In this
paper we compare the features of Bohmian trajectories with those
displayed by the trajectories determined using the momentum
distribution (MD trajectories) associated with the wave function of the
particle. As is well known, Bohmian trajectories follow the flux lines
of the quantum flow, and therefore reproduce exactly the quantum
mechanical distribution in both the far and near field \cite{ref1,%
ref2,ref3,ref4,ref6,ref8}. In
particular, it is remarkable the consistency of a set of Bohmian
trajectories in the near field behind a multiple slit grating within
the context of the Talbot effect \cite{ref8}.
This effect has been observed experimentally with relatively heavy
particles, such as Na atoms \cite{ref8-a} or Bose-Einstein condensates
\cite{ref8-b}.

Moreover, here we also study the momentum distribution associated with
Bohmian trajectories, and compare it with the momentum distribution
associated with the wave function of the particle. We find that in the
far field these two distributions are identical. However, in the near
field, the distribution of transverse momenta associated with the
Bohmian trajectories changes with the distance from the grating, and
therefore differs from the momentum distribution associated with the
wave function.

The essential feature of the Bohmian deterministic trajectories is that
a particle passing through different slits will never reach the same
point on the detection screen \cite{ref4}. On the contrary, this limitation
does not hold for MD trajectories arising from different slits, which
may reach the same point on the screen \cite{ref5,ref7}. The momenta of particles
moving along MD trajectories are distributed according to the momentum
distribution determined by the wave function of the particle. The
probability of arrival of a particle to a given point on the screen
depends on: (1) the slit through which it passed and (2) the presence
of other slits. The second type of dependence is a consequence of the
dependence of this probability on the transverse momentum distribution,
which depends on the properties of the grating as a whole \cite{ref5}. As
shown, MD trajectories reproduce fairly well the quantum mechanical
space distribution in the far field.

\section{Wave function of a particle behind a grating}
\label{sec:2}

\begin{figure*}
 \sidecaption
 \epsfxsize=3.5in {\epsfbox{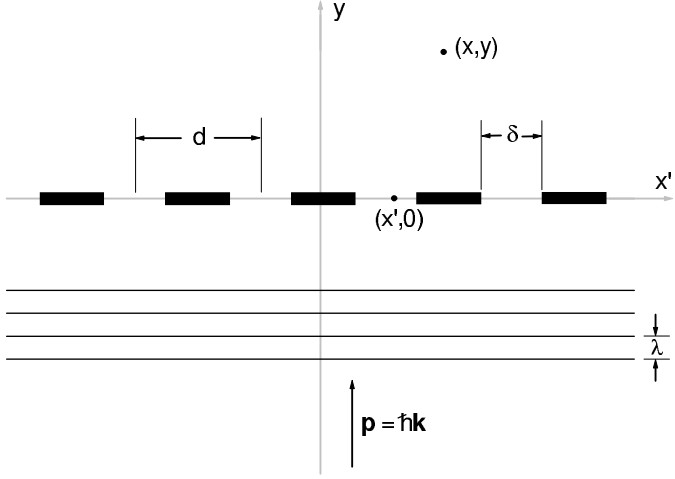}}
 \caption{\label{fig:0} Scheme of the physical system studied.
 Note that $x'$ refers to the coordinate along the grating, while
 $x$ is the actual coordinate out of the grating.}
\end{figure*}

The motion of a particle behind a grating is determined by its wave
function. Here, we assume that in front of the grating ($y<0$) we have
a plane wave with initial momentum $\vec p=\hbar\vec k=\hbar k\vec
i_y$ (and de Broglie wavelength $\lambda = 2\pi/k$), moving along the
longitudinal direction $y$. Behind the grating
($y\ge0$), the solution can be expressed as a product of the
longitudinal part, a plane wave, and the transverse part, i.e.,
\begin{equation}
\Psi(x,y,t)=Be^{iky}e^{-i\omega t}\psi(x,t),\label{eq1}
\end{equation}
where $B$ is a normalization constant. Arsenovi\'c et al.\ have shown
\cite{ref5} that the transverse part (for simplicity taken to be
one-dimensional) is given by
\begin{equation}
\psi(x,t)=\frac{1}{\sqrt{2\pi}}\int_{-\infty}^{+\infty}dk_x\,c(k_x)e^{ik_xx}e^{-i\hbar
k_x^2t/2m},\label{eq2}
\end{equation}
where
\begin{equation}
c(k_x)=\frac{1}{\sqrt{2\pi}}\int_{-\infty}^{\infty}dx\,\psi(x,0)e^{-ik_xx}\label{eq3}
\end{equation}
is the Fourier transform of the initial transverse wave function
$\psi(x,0)$, which is the wave function just behind the grating, is
related to the transmission or window function associated with the
grating, and refers to the time of passage of a particle through the
grating. Substituting (\ref{eq3}) into (\ref{eq2}), $\psi(x,t)$ is
expressed in terms of the initial wave function as
\begin{subequations}
\label{eq4}
\begin{equation}
\psi(x,t)=\frac{\sqrt m}{\sqrt{2\pi\hbar t}} \ \!
e^{-i\pi/4}\int_{-\infty}^\infty\psi(x^\prime,0)e^{im(x-x^\prime)^2/2\hbar
t}dx^\prime.\label{eq4a}
\end{equation}
By assuming that the motion along the $y$-axis is classical, i.e., it
satisfies the relation $y=vt=(\hbar k/m)t$, one finds the dependence of
$\psi(x,t)$ on the longitudinal coordinate $y$:
\begin{equation}
\psi(x,t=ym/\hbar k)=\frac{\sqrt k}{\sqrt{2\pi y}} \ \!
e^{-i\pi/4}\int_{-\infty}^{\infty}\psi(x^\prime,0)e^{ik(x-x^\prime)^2/2y}dx^\prime.\label{eq4b}
\end{equation}
\end{subequations}
Expressions (\ref{eq4a}) and (\ref{eq4b}) are particularly useful when
$\psi(x,0)$ consists of discrete pieces of elements where it is zero.
In the case of a one-dimensional grating with $n$ slits of equal width
$\delta$ (see Fig.~\ref{fig:0}), from the boundary and normalization
conditions follows
\begin{displaymath}
\psi(x,0)=\frac{1}{\sqrt{n\delta}}
\end{displaymath}
at the openings, and
\begin{displaymath}
\psi(x,0)=0
\end{displaymath}
elsewhere (i.e., outside the openings).
Note that this is an idealization of the effect that a perfect periodic
grating with full transmission in the openings would have on a plane
wavefront that would reach it.
Since all the points on this wavefront have the same phase, the
transmitted pieces of wave function described above will also have the
same phase.
We can neglect this phase, since introducing a constant phase factor
will not alter the physics of the problem.
The probability amplitude of the particle transverse momentum $\bar
c(p_x)=c(k_x)/\sqrt\hbar=c(p_x/\hbar)/\sqrt\hbar$ is then given by
\begin{equation}
c(k_x)=\sqrt{\frac{2}{\pi n\delta}}\frac{\sin\left(\frac{k_x\delta}{2}\right)}{k_x}\frac{\sin\left(\frac{k_xdn}{2}\right)}{\sin\left(\frac{k_xd}{2}\right)},\label{eq5}
\end{equation}
where $d$ is the distance between the center of two consecutive slits.

In Optics it is common to consider sharp-edge slits to study
interference and diffraction, as in the model described above.
Changing the shape of the initial wave function would not induce
significant changes in the momentum distribution.
For example, one may use Gaussians instead of square wave functions,
i.e., at the openings we have
\begin{equation}
\psi(x,0) \propto e^{-(x-x_c)^2/a^2} ,
\end{equation}
where $x_c$ is the center of the corresponding opening and $a$ is of
the order of the slit width.
Then a different multiplicative Gaussian factor appears in the first
contribution to the momentum distribution, thus $c(k_x)$ becoming
\begin{equation}
c(k_x) \propto e^{-k_x^2 a^2/4} \ \! \frac{\sin\left(\frac{k_xdn}{2}\right)}{\sin\left(\frac{k_xd}{2}\right)}.\label{eq5-b}
\end{equation}

For $x\gg x^\prime$, the term ${x^\prime}^2$ can be neglected in the
exponential function inside the integral in (\ref{eq4}). By doing that,
one obtains \cite{ref5}
\begin{equation}
\psi(x,t)=\sqrt{\frac{m}{2\pi\hbar t}}e^{-i\pi/4}e^{i\frac{x^2m}{2\hbar
t}}\int_{-\infty}^\infty dx^\prime\,\psi(x^\prime,0)e^{-i\frac{x^\prime
xm}{\hbar t}}.\label{eq6}
\end{equation}
The integral in (\ref{eq6}) is nothing else but the function $c$
defined in (\ref{eq3}), taking into account that the variable $xm/\hbar
t$ plays the role of $k_x$. Therefore, one can write
\begin{equation}
\psi(x,t)=\frac{m}{\hbar t}e^{-i\pi/4}e^{ix^2m/2\hbar
t}c\left(x\frac{m}{\hbar t}\right).\label{eq7}
\end{equation}
The wave function (\ref{eq1}), where the transverse part is given by
(\ref{eq2}) or (\ref{eq4}), explains in a {\em unified way} many
effects and properties of particle diffraction and interference
\cite{ref7}, in particular the Talbot effect. Note that, for a periodic
(infinite) grating, the integral in (\ref{eq2}) reduces to the sum over
discrete momentum values and one can easily see that at integer
multiples of $2L_T$ ($L_T=d^2/\lambda$ being the Talbot distance
\cite{ref8,ref8-a,ref8-b}) the
transverse wave function is simply equal to the wave function on the
grating. The grating pattern is also repeated at odd multiples of
$L_T$, but shifted by half a grating period, $d/2$, along $x$-axis.

In the case of a finite grating, an approximate form of the Talbot
effect appears \cite{ref7,ref9}. Namely, in the near field one finds
self-images distorted at the ends.

\section{Momentum distribution trajectories}
\label{sec:3}

Interference experiments with beams such that only one particle is
launched at a time have demonstrated that particles accumulate with
time on the detection screen, building up an interference pattern after
some time. In order to describe the emergence of the interference
pattern through an accumulation of single particle events one has to
assume that particles move along certain trajectories.

The theoretical investigation of the topology of the trajectories is
based on various assumptions. Arsenovi\'c et al. \cite{ref5} assumed
that trajectories were straight lines starting at different positions
on the slits. The longitudinal momentum of a particle moving along such
a trajectory is equal to the longitudinal momentum of the incident beam
that reaches the grating, and the distribution of the transverse
momentum of the particles is given by $\vert\bar c(p_x)\vert^2$. This
is why we call these trajectories the {\em momentum distribution} (MD)
{\it trajectories}.

Within the MD approach, a particle can start from any slit; if it has
the right transverse momentum, it will reach the chosen detection spot.
The expression for the screen arrival probability based on the idea of
MD trajectories \cite{ref5} reads as
\begin{equation}
\tilde
P(x,t)=\int_{-\infty}^{+\infty}dk_x\int_{-\infty}^{+\infty}dx^\prime\vert
c(k_x)\vert^2\vert\psi(x^\prime,0)\vert^2\delta(x-x^\prime-\hbar k_xt/m).\label{eq8}
\end{equation}
The total probability can be expressed as a sum of $n$ terms
\begin{equation}
\tilde P(x,t)=\sum_{i=1}^n\tilde P_i(x,t),\label{eq9}
\end{equation}
where
\begin{equation}
\tilde P_i(x,t)=\frac{1}{n\delta}\int_{m(x-x_r^i)/\hbar
t}^{m(x-x_l^i)/\hbar t}\vert c(k_x)\vert^2dk_x\label{eq10}
\end{equation}
is the probability that a particle which passed through the $i$-th slit
of the grating arrives to a point $(x,y=\hbar kt/m)$ at time $t$. Here
$x_l^i$ and $x_r^i$ are the coordinates of the left and right edges of
the $i$-th slit.
Note that, as said in the previous Section, the momentum distribution
$c(k_x)$ depends on the initial wave function chosen.
This will only influence the probabilities along the $x$ direction, but
not the shape of the MD trajectories.

Numerical calculations for various gratings show \cite{ref5,ref9,ref10}
that the distributions $\tilde P(x,t)$ and $\vert\psi(x,t)\vert^2$ are
almost identical in the far field. In the near field, both
distributions look qualitatively similar, however they differ
numerically as well as in certain important details. This is
understandable because near the slits the topology of particle
trajectories is more complicated than in the far field, where the
straight-line approximation works fairly well.

\section{Bohmian trajectories}
\label{sec:4}

The Bohmian trajectories associated with the state of a single
particle, $\Psi(\vec r,t)$, are determined \cite{ref1} from the
differential equation ({\em guidance condition})
\begin{equation}
\vec v(\vec r,t)=\frac{d\vec r}{dt}=\frac 1m\nabla S(\vec r,t),\label{eq11}
\end{equation}
where $S(\vec r,t)$ is the phase of the wave function written in the
polar form, i.e.,
\begin{equation}
\Psi(\vec r,t)=Re^{iS/\hbar}.\label{eq12}
\end{equation}
It can be easily shown that
\begin{equation}
\nabla S=\hbar \ \! {\rm Im}\left[ \frac{\nabla\Psi}{\Psi} \right] .
\label{eq13}
\end{equation}
Hence, according to (\ref{eq1}), in our case we have
\begin{equation}
\vec v=\frac{d\vec r}{dt}=\frac\hbar m
\left\{\vec i_yk+\vec i_x{\rm Im}
 \left[\frac{\partial_x\psi(x,t)}{\psi(x,t)}\right]\right\} ,
  \label{eq14}
\end{equation}
where $\partial_x = \partial/\partial x$.
The equation of motion along the $y$-axis,
\begin{equation}
v_y=\frac{dy}{dt}=\frac{\hbar k}{m},\label{eq15}
\end{equation}
is very simple, and has the solution
\begin{equation}
y=y_0+\frac{\hbar k}mt.\label{eq16}
\end{equation}
On the other hand, the equation of motion along the $x$-axis
\begin{equation}
v_x=\frac{dx}{dt}=\frac\hbar m \ \! {\rm Im}\left[
\frac{\partial_x\psi(x,t)}{\psi(x,t)} \right]
 \label{eq17}
\end{equation}
can not be solved analytically for all values of $x$.
Unlike MD trajectories, for Bohmian trajectories both their
distribution and shape are influenced by the choice of
the initial wave function \cite{ref3}, although fundamental features
still remain.

\begin{figure*}
 \sidecaption
 \epsfxsize=3in {\epsfbox{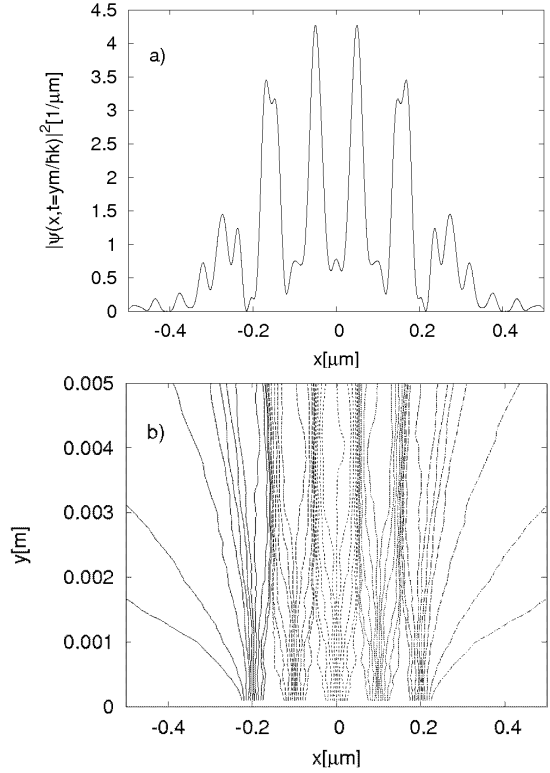}}
 \caption{\label{fig:1} a) Particle distribution function behind a
grating with $n$=5
slits for $y$=1.25$L_T$, where $L_T$ is the Talbot distance
$L_T$=$d^2/\lambda$. b) Trajectories behind a grating with $n$=5 slits
for $y \le 1.25L_T$.
Trajectories passing through different openings are represented with
different line styles.
The values of the parameters used in the calculations are:
$d=0.1$$\times$10$^{-6}$~m, $\delta=0.05$$\times$10$^{-6}$~m,
$m=1.19$$\times$10$^{-24}$~kg, $v=220$~m/s and
$\lambda=2.53$$\times$10$^{-12}$~m.}
\end{figure*}

Equation (\ref{eq17}) can only be
solved analytically in the far field, using the approximate expression
(\ref{eq7}) for $\psi(x,t)$. Using this approximation, Eq.~(\ref{eq17})
reduces to
\begin{equation}
\frac{dx}{dt}=\frac xt,\label{eq18}
\end{equation}
whose solution reads
\begin{equation}
x=\frac{x_0}{t_0} \ \! t.\label{eq19}
\end{equation}
We have found numerical solutions of Eq.~(\ref{eq17}) and plotted the
Bohmian trajectories for particles behind gratings with $n=5$ and
$n=30$ slits. As can be seen in Figs.~1 and 2, the density of Bohmian
trajectories at a certain distance from the grating is in good
agreement with the quantum mechanical probability density,
$\vert\psi(x,t=ym/\hbar k)\vert^2$, in both the far field (see Fig.~2)
and the near field (see Fig.~1). Bohmian trajectories, grouped in
bunches, follow the directions that end at the regions of the different
intensity peaks. Such an agreement was found previously by Sanz et
al.~\cite{ref1,ref3,ref4}, who plotted together the intensity pattern
obtained by means of the standard quantum mechanics and the histogram
obtained by counting Bohmian trajectories (see Fig.~5 in \cite{ref4},
for instance).

The consistency of the set of Bohmian trajectories in the near field
within the context of the Talbot effect found by Sanz and Miret-Art\'es
\cite{ref8} is remarkable. This consistency is also seen in Fig.~3,
which shows the Bohmian trajectories behind a grating with $n=30$ slits
(only one half of the full pattern is shown; the other half is its
mirror image).

\begin{figure*}
 \sidecaption
 \epsfxsize=3in {\epsfbox{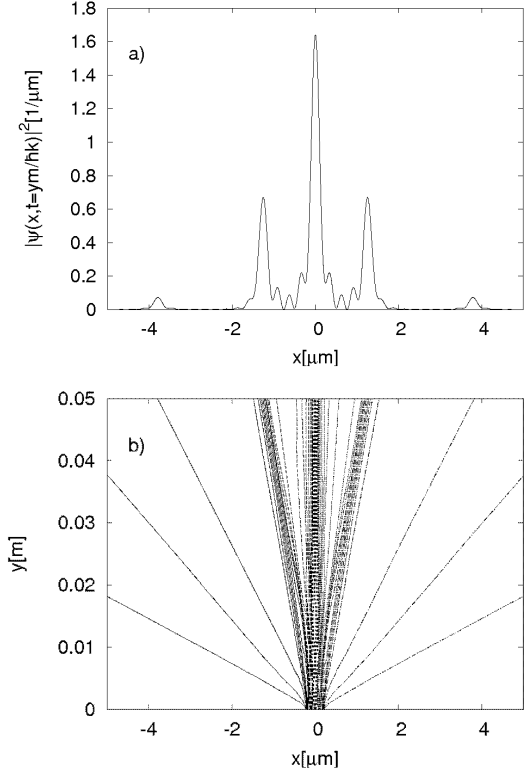}}
 \caption{\label{fig:2} a) Particle distribution function behind a
grating with $n$=5
slits for $y$=12.5$L_T$, where $L_T$ is the Talbot distance
$L_T$=$d^2/\lambda$. b) Trajectories behind a grating with $n$=5 slits
for $y \le 12.5L_T$.
The values of the parameters used in the calculations are:
$d=0.1$$\times$10$^{-6}$~m, $\delta=0.05$$\times$10$^{-6}$~m,
$m=1.19$$\times$10$^{-24}$~kg, $v=220$~m/s and
$\lambda=2.53$$\times$10$^{-12}$~m.}
\end{figure*}

\begin{figure*}
 \begin{center}
 \epsfxsize=5in {\epsfbox{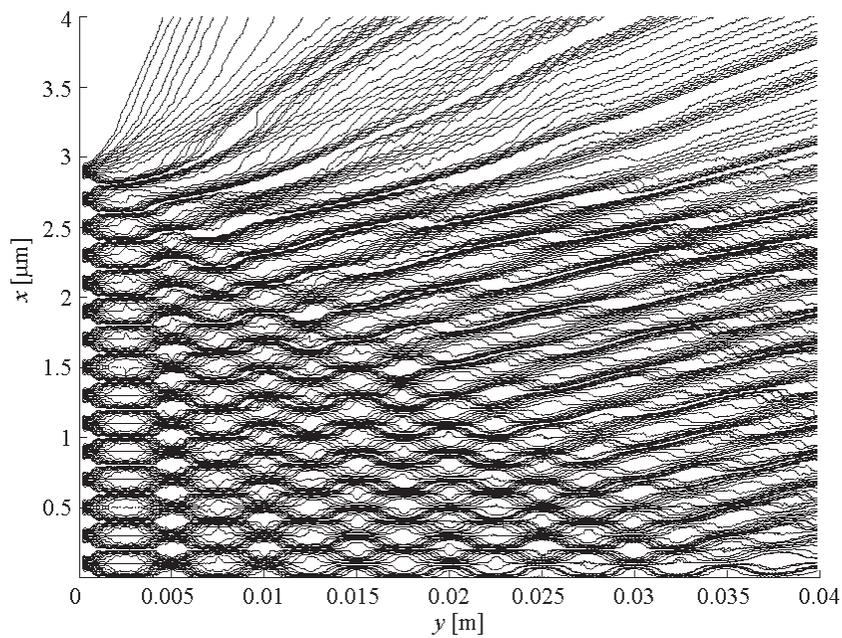}}
 \end{center}
 \caption{\label{fig:3} Bohmian trajectories behind one half of a Ronchi grating with
$n$=30 slits. The values of the parameters used in the calculations are:
$d=0.2$$\times$10$^{-6}$~m, $\delta=0.1$$\times$10$^{-6}$~m,
$k=(\pi/8)$$\times$10$^{12}$~m$^{-1}$, $m=3.8189$$\times$10$^{-26}$~kg
and $v=1084$~m/s.}
\end{figure*}

\section{Comparison between Bohmian trajectories and MD trajectories}
\label{sec:5}

An essential feature of the Bohmian deterministic trajectories is that
a particle passing through different slits will not reach the same
point on the detection screen. In the far field, Bohmian trajectories
asymptotically approach straight lines that connect the center of the
grating with the detection spot (see Fig.~4). On the contrary, MD
trajectories from different slits may reach the same point on the
screen, because this trajectories are associate with different values
of transverse momentum. Namely, a particle which leaves certain point
at the slit may have various values of momentum, in accordance with the
transverse momentum distribution.

\begin{figure*}
 \sidecaption
 \epsfxsize=2.5in {\epsfbox{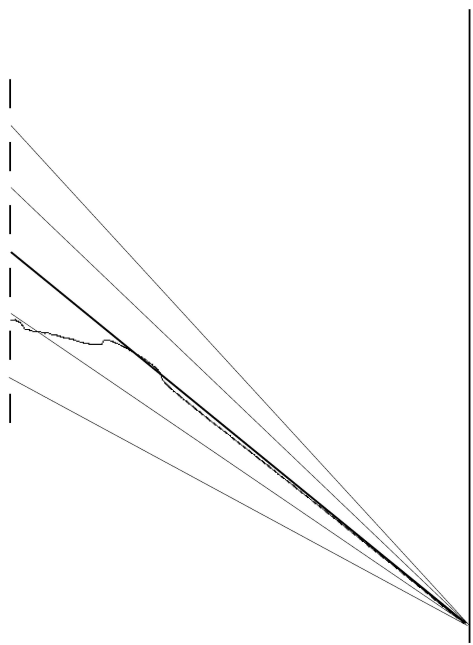}}
 \caption{\label{fig:4} Sketch of MD (five straight lines) and Bohmian (curve) trajectories where
there essential features are compared.
The central among these
straight lines is the asymptote of a Bohmian trajectory.}
\end{figure*}

\section{Momentum distribution}
\label{sec:6}

It is evident that the probability amplitude of the transverse momenta
is an essential feature of the wave function and wave field behind a
grating. Thus, the following question arises: is the distribution of
transverse momenta associated with Bohmian trajectories identical or
different from the distribution $\vert\bar c(p_x)\vert^2=\vert
c(k_x)\vert^2/\hbar$, where $c(k_x)$ is determined by (\ref{eq3})?

Long time ago, based on a general expression for the distribution of
Bohmian momenta, Takabayasi \cite{ref11} concluded that the
aforementioned two distributions were different functions. We have
found that this conclusion is not always true. Using the
Eq.~(\ref{eq17}) for Bohmian trajectories, and the approximation
(\ref{eq7}) for the wave function, we are going to show that the
distributions determined from the wave function and from Bohmian
trajectories are identical in the far field. In the near field, the
distribution of transverse momenta associated with Bohmian trajectories
changes with the distance from the grating and is different from the
distribution $\vert\bar c(p_x)\vert^2$.

\begin{figure*}
 \begin{center}
\resizebox{1.0\columnwidth}{!}{%
  \includegraphics{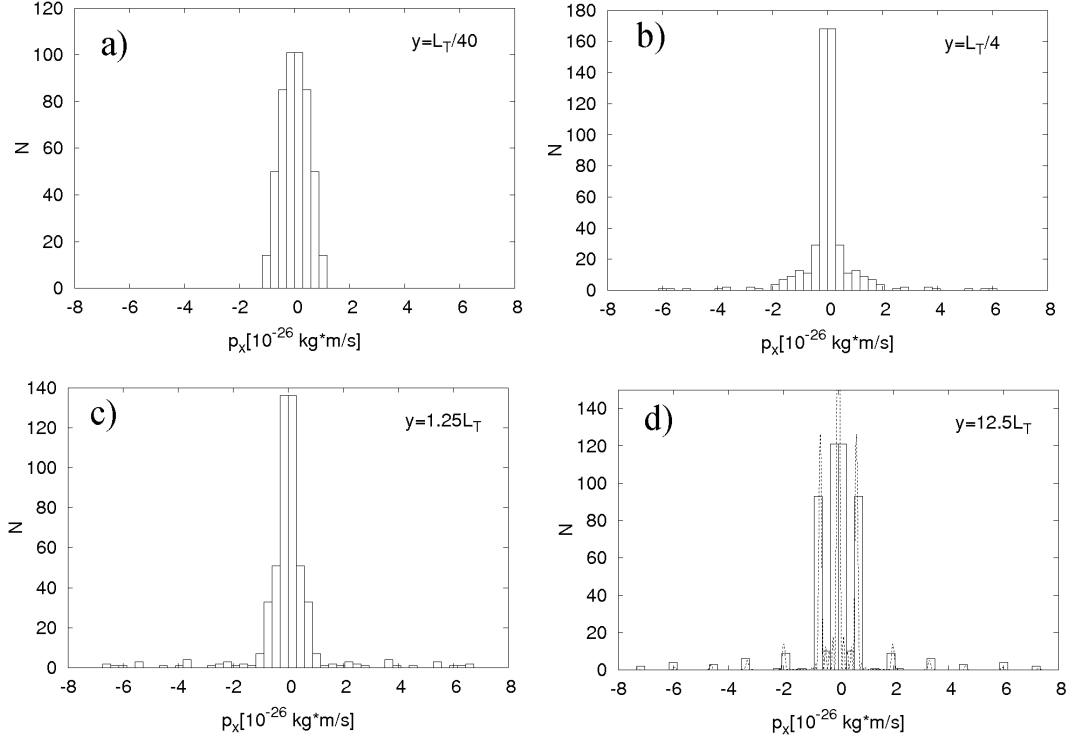} }
 \end{center}
 \caption{\label{fig:5} Histogram of the Bohmian momentum distribution at four
distances from a grating with $n$=5 slits: (a) $y$=$L_T/40$,
(b) $y$=$L_T/4$, (c) $y$=1.25$L_T$ and (d) $y$=12.5$L_T$.
In panel (d), the quantum momentum distribution has also been plotted
with dashed line.}
\end{figure*}

\begin{figure*}
 \begin{center}
\resizebox{1.0\columnwidth}{!}{%
  \includegraphics{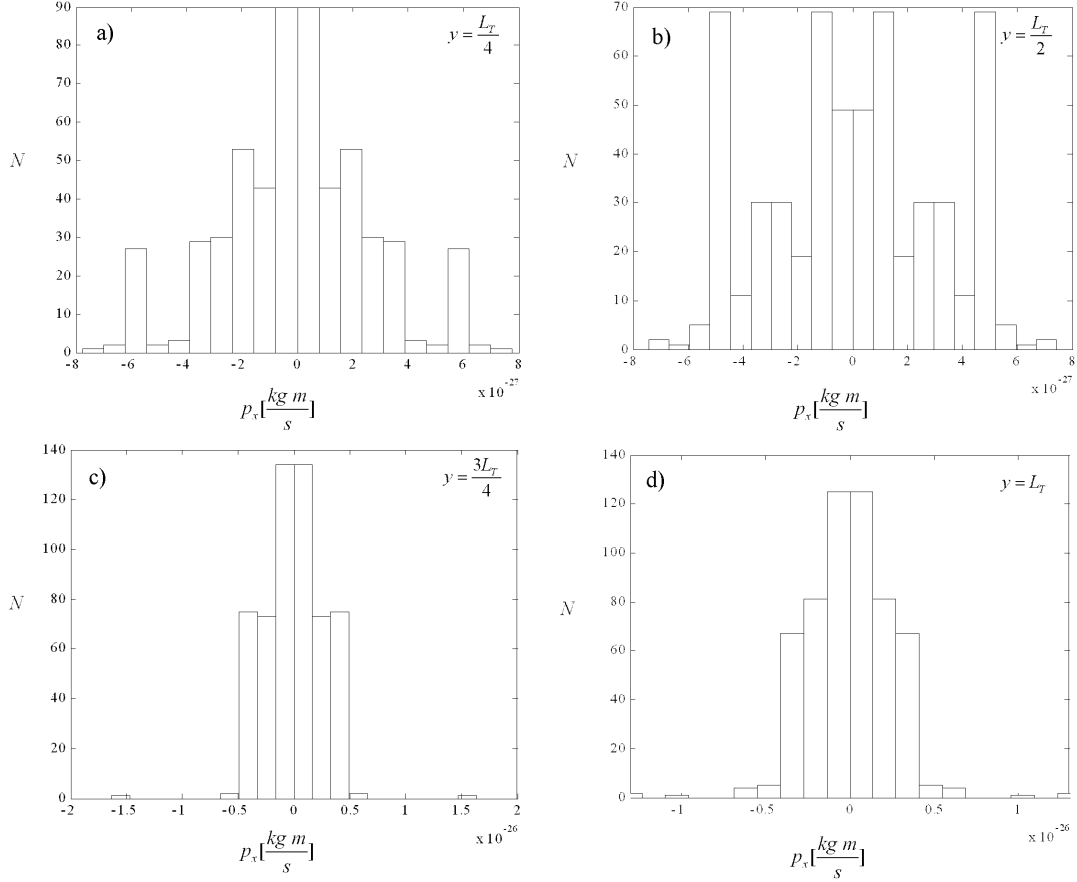} }
 \end{center}
 \caption{\label{fig:6} Histogram of the Bohmian momentum distribution at four
distances from a grating with $n$=30 slits: (a) $y$=$L_T/4$,
(b) $y$=$L_T/2$, (c) $y$=0.75$L_T$ and (d) $y$=$L_T$.}
\end{figure*}

An analytical expression for the transverse momentum distribution in
the far field can be easily obtained. One starts from a general
relation between the probability distribution in the $x$-space and in
the $p_x$-space,
\begin{equation}
P(x,t)dx=\Pi(p_x,t)dp_x.\label{eq20}
\end{equation}
Taking into account that $P(x,t)=\vert\psi(x,t)\vert^2$ as well as
relation (\ref{eq17}), we find the following expression for the
probability distribution of Bohmian momenta along the $x$-axis:
\begin{equation}
\Pi(p_x,t)=\frac{\vert\psi(x,t)\vert^2}{\partial_x p_x}=
\frac{\vert\psi(x,t)\vert^2}{\partial_{xx}S} , \label{eq21}
\end{equation}
where $\partial_{xx} = \partial^2/\partial x^2$.
Using the approximate expression (\ref{eq7}) for the transverse wave
function, which is valid in the far field, we find:
\begin{eqnarray}
\vert\psi(x,t)\vert^2&=&\frac{m}{\hbar t}\left\vert c\left(x\frac m{\hbar
t}\right)\right\vert^2\label{eq22}\\
\frac{\partial^2S}{\partial x^2}& =&\frac mt.\label{eq23}
\end{eqnarray}
Therefore, the probability distribution of Bohmian momenta in the far
field will be
\begin{equation}
\Pi(p_x,t)=\frac1\hbar\vert c(k_x)\vert^2=\vert\bar c(p_x)\vert^2.\label{eq24}
\end{equation}
We do not have an analytic expression for $\Pi(p_x,t)$ in the near
field, but the histograms of the probability density of Bohmian momenta
plotted in Figs.~5 and 6 show that it changes with the distance from
the grating. This means that, in the near field, the probability
density of Bohmian momenta differs from the momentum distribution
$\vert\bar c(p_x)\vert^2$.

\section{Conclusions}
\label{sec:7}

By appealing to descriptions in terms of quantum particle trajectories
in interference experiments, one can understand how interference
patterns emerge from the accumulation of single particle events.
This justifies theoretical
investigations of properties and statistics of particle trajectories.

In Bohmian mechanics trajectories reproduce exactly the quantum
mechanical space distribution in both the far and near fields. The
consistency of the set of Bohmian trajectories in the near field within
the context of the Talbot effect, demonstrated by Sanz and Miret-Art\'{e}s
\cite{ref8}, is remarkable.

The probability amplitude of transverse momenta is also an essential
feature of the wave function and the wave field behind the grating.
Taking this into account, the following question arises: should
particle trajectories comply with the transverse momentum distribution?
From our numerical calculations and analytical treatments, it follows
that the distributions of transverse momentum determined from the wave
function and from Bohmian trajectories are identical in the far field.
On the other hand, in the near field, the distribution of transverse
momenta associated with the Bohmian trajectories changes with the
distance from the grating and is different from the distribution
$\vert\bar c(p_x)\vert^2$.

Considering that the answer to the aforementioned question should be
positive, Arsenovi\'c et al. proposed \cite{ref5} to approximate
trajectories by straight lines and to assume that the distribution of
particle momenta is determined by the wave function. These
trajectories, which we call MD trajectories, reproduce fairly well the
quantum mechanical space distribution in the far field
\cite{ref5,ref9}. However, in the near field the agreement is not so
satisfactory. It seems that a better agreement of the space
distribution derived from MD trajectories with the quantum mechanical
space distribution in the near field could be obtained by combining
peaces of various Bohmian trajectories \cite{ref1,ref2,ref3,ref4,ref8},
and by studying also lines of a quantum mechanical current.

\begin{acknowledgement}
Davidovi\'c, Arsenovi\'c and Bo\v zi\'c acknowledge support from the
Ministry of Science of Serbia under Project ``Quantum and Optical
Interferometry'', N 141003; and Sanz and Miret-Art\'es acknowledge
support from the DGCYT (Spain) under Project FIS2004-02461.
A.S. Sanz would also like to thank the Spanish Ministry of Education
and Science for a ``Juan de la Cierva'' Contract.
\end{acknowledgement}


\begin{thebibliography}{}

\bibitem{ref1}
A.S. Sanz, F. Borondo, S. Miret-Art\'es, Phys. Rev. B \textbf{61},
7743 (2000)

\bibitem{ref2}
A.S. Sanz, F. Borondo, S. Miret-Art\'es, Europhys. Lett. \textbf{55},
303 (2001)

\bibitem{ref3}
A.S. Sanz, F. Borondo, S. Miret-Art\'es, J. Phys.: Condens. Matter
\textbf{14}, 6109 (2002)

\bibitem{ref4}
R. Guantes, A.S. Sanz, J. Margalef-Roig, S. Miret-Art\'es, Surf.
Sci. Rep. \textbf{53}, 199 (2004)

\bibitem{ref5}
D. Arsenovi\'c, M. Bo\v zi\'c, L. Vu\v skovi\'c, J. Opt. B: Quantum
Semiclassical Opt. \textbf{4}, S358 (2002)

\bibitem{ref6}
M. Gondran, A. Gondran, Am. J. Phys. \textbf{73}, 507 (2005)

\bibitem{ref7}
M. Bo\v zi\'c, D. Arsenovi\'c, Acta Physica Hungarica B
\textbf{26/1-2}, 219 (2006)

\bibitem{ref8}
A.S. Sanz, S. Miret-Art\'es, J. Chem. Phys. \textbf{126}, 234106 (2007)

\bibitem{ref8-a}
M.S. Chapman, C.R. Ekstrom, T.D. Hammond, J. Schmiedmayer,
B.E. Tannian, S. Wehinger, D.E. Pritchard, Phys. Rev. A
{\bf 51} R14 (1995)

\bibitem{ref8-b}
L. Deng, E.W. Hagley, J. Denschlag, J.E. Simsarian, M. Edwards,
C.W. Clark, K. Helmerson, S.L. Rolston, W.D. Phillips,
Phys. Rev. Lett. {\bf 83}, 5407 (1999)

\bibitem{ref9}
M. Bo\v zi\'c, D. Arsenovi\'c, L. Vu\v skovi\'c, Phys. Rev. A
\textbf{69}, 053618 (2004)

\bibitem{ref10}
M. Bo\v zi\'c, D. Arsenovi\'c, L. Vu\v skovi\'c, Concepts of Physics
\textbf{I{}I}, 163 (2005)

\bibitem{ref11}
T. Takabayasi, Prog. Theor. Phys. \textbf{8}, 143 (1952)

\end{thebibliography}
\end{document}